\documentclass[11pt]{article}
\usepackage{amsmath,amssymb}
\usepackage[nospace]{cite}
\renewcommand{\d}{\mathrm{d}}
\newcommand{\e}{\mathrm{e}}
\renewcommand{\i}{\mathrm{i}}
\newcommand{\qe}{ Q_\mathrm{e}}
\newcommand{\qm}{ Q_\mathrm{m}}
\newcommand{\qem}{ Q_\mathrm{e}\!\!\cdot \! Q_\mathrm{m}}
\newcommand{\Qe}{Q_\mathrm{e}}
\newcommand{\Qm}{Q_\mathrm{m}}
\newcommand{\Qem}{Q_\mathrm{e}\!\!\cdot \! Q_\mathrm{m}}

\begin{document}
\begin{titlepage}
\begin{center}
\hfill LMU-TPS-04/14 \\
\hfill ITP-UU-04/44  \\
\hfill SPIN-04/26   \\
\hfill AEI-2004-121 \\
\hfill FSU-TPI 10/04 \\
\hfill {\tt hep-th/0412287}\\
\vskip 20mm

{\Large \textbf{ Asymptotic degeneracy of dyonic N=4 string states \\[2mm] 
and black hole entropy
}}
\vskip 10mm

\textbf{G.~L.~Cardoso$^{a}$, B. de Wit$^b$, J.~K\"appeli$^{c}$, 
       and  T. Mohaupt$^{d}$}

\vskip 4mm
$^a${\em Department f\"ur Physik, 
Ludwig-Maximilians-Universit\"at M\"unchen, Munich, Germany}\\
{\tt gcardoso@theorie.physik.uni-muenchen.de}\\[2mm]
$^b${\em Institute for Theoretical Physics} and {\em Spinoza
  Institute,\\ Utrecht University, Utrecht, The Netherlands}\\
{\tt  B.deWit@phys.uu.nl} \\[2mm]
$^c${\em Max-Planck-Institut f\"ur Gravitationsphysik,
  Albert-Einstein-Institut, Potsdam, Germany}\\
{\tt kaeppeli@aei.mpg.de}\\[2mm]
$^d${\em Theoretisch-Physikalisches Institut, Friedrich-Schiller
  Universit\"at Jena, Jena, Germany}\\
{\tt moh@tpi.uni-jena.de}

\vskip 6mm
\end{center}
\vskip .2in
\begin{center} {\bf ABSTRACT } \end{center}
\begin{quotation}\noindent
  It is shown that the asymptotic growth of the microscopic degeneracy of 
  BPS dyons in four-dimensional $N=4$ string theory captures the known 
  corrections to the macroscopic entropy of four-dimensional
  extremal black holes. These corrections are subleading in the limit
  of large charges and originate both from the presence of interactions in the
  effective action quadratic in the Riemann tensor and from 
  non-holomorphic terms. The presence of the non-holomorphic 
  corrections and their contribution to the thermodynamic free energy is
  discussed. It is pointed out that the expression for the macroscopic
  entropy, written as a function of the dilaton field, is stationary at the
  horizon by virtue of the attractor equations.
\end{quotation}

\vfill
\end{titlepage}

\eject
\section{Introduction}
\setcounter{equation}{0}
String theory predicts deviations from the Bekenstein-Hawking area law for the
entropy of black holes. For large charges the microstate counting yields a
statistical entropy which generically coincides with the area of the
corresponding macroscopic black hole solutions \cite{Strominger:1996sh}. In a number
of cases, also subleading corrections can be determined. This is especially
relevant for the heterotic string, where certain elementary string states can
be identified with black holes. For 1/2-BPS states the charges are 
restricted and as a result the leading contributions to the entropy
vanish \cite{Sen:1995in}. Therefore, the dominant contribution to area and entropy will come
from subleading terms, which do not obey the proportionality relation as
implied by the area law. In all of this it is important that one is dealing
with BPS states, corresponding to extremal black holes, so that the effects of
string interactions remain under control. 

On the macroscopic side subleading corrections have been extensively
studied in the framework of four-dimensional $N=2$ supergravity, where they
are induced by couplings in the effective Wilsonian action that are quadratic
in the curvature tensor. In \cite{LopesCardoso:1998wt} a general formula for
the entropy was presented in terms of the homogeneous, holomorphic function
$F(X,A)$, in which these $N=2$ supergravity Lagrangians are encoded. Here
the holomorphic variables $X$ are related to the complex moduli associated with
vector supermultiplets, and the dependence on $A$ characterizes the terms
quadratic in the Riemann tensor. The $N=2$ entropy formula was successfully
confronted with the results from microstate counting for a class of black
holes arising in compactifications of M-theory and type-IIA string theory
\cite{Maldacena:1997de,Vafa:1997gr}. In the M-theory setting the microscopic
object that enters is the five-brane, wrapped on a four-cycle of a
Calabi-Yau three-fold. An important role in the supergravity 
analysis is played by the attractor equations
\cite{Ferrara:1995ih,Strominger:1996kf, Ferrara:1996dd} which
fix the values of the moduli at the horizon in terms of the black hole
charges. For effective actions with interactions quadratic in the
curvature the validity of these attractor equations was established in
\cite{LopesCardoso:2000qm}.

Recently, it was shown \cite{Ooguri:2004zv} that the $N=2$ entropy
formula can be rewritten as a Legendre transform of a real function,
${\cal F}(\Qm,\phi)$, where $\phi$ denotes the electric potentials at
the horizon and the electric charges are given by $\Qe= \partial{\cal
  F}/\partial\phi$. 
Subsequently ${\cal F}$ was identified with the logarithm of a mixed black
hole partition function, which is microcanonical with respect to the magnetic
charges $\Qm$ and canonical with respect to the electric potentials $\phi$ at 
the horizon. It was then conjectured that this partition function can
be written as a Laplace transform of the microscopic black hole
degeneracies $d(\Qe,\Qm)$. The original result of \cite{LopesCardoso:1998wt}
is to be recovered in the limit of large electric charges. 
These observations have rekindled the interest in the question of how the
entropy formula is precisely related to the actual microscopic degeneracies. 
In this paper we will study the relation between the entropy formula and
the microscopic degeneracies for $N=4$ dyons proposed in
\cite{Dijkgraaf:1996it} beyond the leading order. 

In \cite{LopesCardoso:1999ur} the modified entropy formula was already applied
to heterotic black holes. Although the formula was initially derived for $N=2$ 
supergravity, the result can readily be generalized to the case of heterotic
$N=4$ supersymmetric compactifications. This involves an extension of the
target-space duality group from ${\rm SO}(2,18)$ to ${\rm SO}(6,22)$ with a
corresponding extension of the charges and the moduli. The $N=4$
supersymmetric heterotic models have dual realizations as type-II string
compactifications on $K3\times T^2$. In contrast to $N=2$ Calabi-Yau
compactifications, the holomorphic function which encodes the effective
Wilsonian action is severely restricted in the $N=4$ case.  Therefore it
is often possible to obtain exact predictions in this context. In
\cite{LopesCardoso:1999ur} the perturbative holomorphic function for the
$N=4$ heterotic theory was appropriately extended in order to obtain results
that were invariant under both target-space duality and $S$-duality. While the
first 
requirement posed no particular problems, the latter necessitates the addition
of non-holomorphic terms. This feature is not unexpected: the Wilsonian
couplings are holomorphic but may not fully reflect the symmetries of the
underlying theory, while the physical couplings must reflect the symmetry and
may thus have different analyticity properties. It turned out that the
non-holomorphic terms are determined uniquely by requiring $S$-duality and
consistency with string perturbation theory, and are in accord with the
results of \cite{Harvey:1996ir}.

Including the non-holomorphic corrections, the result of
\cite{LopesCardoso:1999ur} can be summarized as
follows. The non-trivial attractor equations are the ones that determine the
horizon value of the complex dilaton field $S$ in terms of the black hole 
charges. They read as follows,  
\begin{eqnarray}
\label{eq:nonholostab}
\vert S\vert^2 \, \Qm^2 &=& \Qe^2 + \frac{ 128\,c_1}{\pi} ( S + {\bar S}) \, 
\Big(S \frac{\partial}{\partial S}+ \bar S \frac{\partial}{\partial \bar
  S}\Big) 
 \log\left[ (S+\bar S)^6\vert\eta(S)\vert^{24}\right] \;, \nonumber\\ 
(S - {\bar S}) \,\Qm^2 &=& {}-2 \,\i \, \Qem - \frac{128\, c_1}{\pi} (S + {\bar
  S}) \, \Big(\frac{\partial}{\partial S} - \frac{\partial}{\partial\bar
  S}\Big) 
\log\left[ (S+\bar S)^6\vert \eta(S)\vert^{24}\right] \;.
\end{eqnarray}
Here $\Qe^2$, $\Qm^2$ and $\Qem$ are the three target-space duality
invariant contractions of the electric and the magnetic charges, $\Qe$
and $\Qm$, which, in the $N=4$ extension, take values in the
lattice $\Gamma^{6,22}$. From the supergravity calculations there is
no intrinsic definition of the lattice of charges and consequently the dilaton
normalization is a priori not known.\footnote{
 We have deviated from the notation in
 \cite{LopesCardoso:1999ur} and employ the definitions: $\Qe^2=
 -\langle M,M\rangle$, $\Qm^2= - \langle N,N\rangle$ and  $\Qem= M\cdot
 N$. Note that, in the limit of large charges, $\Qe^2$ and $\Qm^2$ are 
 negative.  
}  
Our definition of the Dedekind eta-function $\eta(S)$
follows from the asymptotic formula, $\log \eta(S)\approx -\frac1{12}
\pi\,S + {\mathrm e}^{-2\pi S}+ {\cal O}({\mathrm e}^{-4\pi S})$. We also
recall that $\eta^{24}(S)$ is a modular form of degree 12, so that   
$\eta^{24}(S^\prime) = (\i \, c\,S+d)^{12}\,\eta^{24}(S)$, where $S^\prime$ is
the transformed dilaton field, given below in (\ref{eq:S-duality}). The
constant $c_1$ must be equal to $c_1=-\frac1{64}$, as we shall discuss later.

The expression for the macroscopic entropy reads, 
\begin{eqnarray}
  \mathcal{S}_\text{macro} = 
- \pi \left[ \frac{\Qe^2 - \i \Qem \, (S - {\bar S}) + \Qm^2 \,|S|^2} 
{S + {\bar S}} \right] + 128\,c_1\, \log\left[ (S + {\bar S})^6
  |\eta(S)|^{24}\right] \;, 
\label{eq:nonholoentropy}
\end{eqnarray}  
with the dilaton subject to (\ref{eq:nonholostab}). The first term in this
equation corresponds to one-fourth of the horizon area, 
which, via (\ref{eq:nonholostab}), is affected by the various corrections. 
The second term represents an extra modification, which explicitly contains
the non-holomorphic correction. The above results are invariant under
target-space duality and $S$-duality. As 
explained above, this was achieved at the price of including 
non-holomorphic terms, here residing in the $\log(S+\bar S)$ terms, at an
intermediate stage of the calculation. Under $S$-duality the dilaton field
transforms in the usual manner under ${\rm SL}(2,\mathbb{Z})$ and the
${\rm SO}(6,22)$ invariant contractions of the charges transform according to
the corresponding arithmetic subgroup of ${\mathrm{SO}}(2,1)$, 
\begin{eqnarray}
  \label{eq:S-duality}
   S &\to& \frac{a\,S -\i b}{\i c\,S +d}\,,\nonumber\\
  \Qe^2 &\to& a^2\,\Qe^2 +b^2\,\Qm^2 + 2\,ab\,\Qem\,, \nonumber\\
  \Qm^2 &\to& c^2 \Qe^2 + d^2 \,\Qm^2 + 2\,cd\,\Qem\,, \nonumber\\  
  \Qem &\to&  ac\, \Qe^2 + bd\,\Qm^2 +(ad+bc) \,\Qem \,.
\end{eqnarray}
Here $a,b,c,d$ are integer-valued with $ad-bc=1$, such that they preserve the
charge lattice. Observe that the above transformation rules then fix 
the normalization of the dilaton field. 

In string perturbation theory the real part of $S$ becomes large and
positive, and one can neglect the exponential terms of the Dedekind
eta-function. In that approximation the imaginary part of $S$ equals
${\rm Im}\,S= - {Q_\mathrm{e}\!\!\cdot \! Q_\mathrm{m}}/\Qm^2$ and the real part is determined by a quadratic
equation,  
\begin{equation}
  \label{eq:re-S}
  \tfrac1{4} \Qm^2 \left(\Qm^2 + 512\,c_1\right)
\, (S+\bar S)^2 -  \frac{768\,c_1}{\pi}\,\Qm^2  
  \,(S+\bar S) = {\Qe^2\,\Qm^2 -(\Qem)^2} \,.
\end{equation}
Obviously, these perturbative results are affected by the presence of the
non-holomorphic corrections. 
Using (\ref{eq:re-S}), we find the following expression for the corresponding
entropy,  
\begin{equation}
\label{eq:entcor}
  \mathcal{S}_{\text{macro}} = - 
{2\pi}  \, \frac {\Qe^2 \,\Qm^2 - (\Qem)^2}{\Qm^2\,(S+\bar S)} 
 + 768\, c_1 \left[\log (S+\bar S) - 1\right] \,.
\end{equation}
In this paper we will be considering the limit of
large charges, where $\Qe^2\, \Qm^2 -(\Qem)^2\gg 1$ and $\Qe^2+\Qm^2$ is large
and negative. We will consider a uniform scaling of all the charges. The
dilaton field will remain finite in that limit; to ensure that it is
nevertheless large, one must assume that $\vert\Qm^2\vert$ is sufficiently
small as compared to $\sqrt{\Qe^2\, \Qm^2 -(\Qem)^2}$.  

In the $N=4$ setting the purely electric or magnetic configurations
constitute 1/2-BPS states, whereas the dyonic ones are 1/4-BPS states.
In the $N=2$ truncation this distinction disappears. While the results of
\cite{LopesCardoso:1999ur} apply to both cases, the purely electric case was
not given much attention at the time. To describe the 1/2-BPS states, we 
assume $\Qe^2$ to be large and negative and $\Qm^2=\Qem=0$ so that we are in a
different domain of the charge lattice, where $\Qe^2 \Qm^2 -(\Qem)^2=0$. 
In this case the leading contributions to the entropy and the area will
vanish. Consequently, the subleading contributions will now dominate and one
finds, 
\begin{eqnarray}
\label{eq:entcor-electric}
  S+\bar S &\approx &\sqrt{\Qe^2/(128\,c_1)}\,,\nonumber \\ 
  \mathcal{S}_{\text{macro}} &\approx& 
2\,\pi\, \sqrt{128\,c_1\, \Qe^2} + 384 \, c_1 \log{\vert\Qe^2\vert}  \,. 
\end{eqnarray}
The leading term in the entropy is now one-half of the area. This
is due to the fact that the terms proportional to the square of the curvature
contribute to the area and entropy with a relative factor 2, as was already
noted in \cite{LopesCardoso:1999ur}, while the `classical' contribution
vanishes. The physical implications of this phenomenon have recently been
discussed in a number of
papers~\cite{Dabholkar:2004yr,Dabholkar:2004dq,Sen:2004dp,Hubeny:2004ji}.   

Let us compare (\ref{eq:entcor-electric}) to the asymptotic degeneracy of
 1/2-BPS states of heterotic string theory, which is given by 
\begin{equation}
  \label{eq:het-string}
  d(\Qe) = \oint {\mathrm d}\sigma\, \frac{{\mathrm
  e}^{\i\pi\sigma\Qe^2}}{\eta^{24}(\sigma)} \approx 
  \exp\left(4\pi\,\sqrt{ \frac{|\Qe^2|}{2}} - \frac{27}{4}
      \log \vert\Qe^2\vert\right)  \,,
\end{equation}
where the integration contour encircles the point $q\equiv\exp(2\pi\i \sigma)
=0$. We therefore find agreement at leading order in large $\vert\Qe^2\vert$,
provided that 
$c_1=-\tfrac1{64}$ as claimed, while the logarithmic corrections fail to
agree. The value for $c_1$ can also be deduced from
string-string duality. For type-II string theory compactified on $K3\times
T^2$, $c_1$ is equal to $-\tfrac1{24\cdot 64}\,\chi$, where $\chi$ is the
Euler number of $K3$. As the latter is equal to 24, one obtains the same value
for $c_1$. It is worth pointing out that the coefficient of the logarithmic
terms (\ref{eq:entcor-electric}) is thus equal to $-6$. The difference with
the corresponding term in (\ref{eq:het-string}) is precisely the contribution
that one obtains from the Gaussian integral when deriving the right-hand side
of (\ref{eq:het-string}) by a saddle-point approximation. It seems unlikely
that this is a coincidence. A recent, extended discussion of this discrepancy
can be found in~\cite{Sen:2004dp}. 

As stated above, this paper will deal with the comparison between the
degeneracy of dyons proposed in \cite{Dijkgraaf:1996it} and the subleading 
corrections that are known from the supergravity description
\cite{LopesCardoso:1998wt,LopesCardoso:1999ur}. One of the main results is
that, in the limit of large charges, the
degeneracy formula leads to precisely the equations (\ref{eq:nonholostab}) and
(\ref{eq:nonholoentropy}). The  paper is organized as follows. In section~2 we 
introduce the formula for the microscopic dyon degeneracies in $N=4$ string
theory. In section~3 we evaluate its asymptotic behaviour in the limit of
large charges and show that it is in precise agreement 
with the results from the macroscopic description. Finally, in section~4, we
present our conclusions. We also discuss the contribution of non-holomorphic
terms to the thermodynamic free energy and some aspects related to the 1/2-BPS
states. Finally we point out that the attractor equations
(\ref{eq:nonholostab}) ensure the stationarity of the expression for the
macroscopic entropy (\ref{eq:nonholoentropy}), written as a function of the
dilaton field. 

\section{Counting dyon states}
Quite some time ago, Dijkgraaf, Verlinde and Verlinde proposed a formula for
the microscopic degeneracies of dyonic states of $N=4$ string theory
\cite{Dijkgraaf:1996it}. The degeneracy is expressed in terms of an integral 
over an appropriate 3-cycle that involves an automorphic form
$\Phi_{10}(\Omega)$,  
\begin{equation}
  \label{eq:dvvdeg}
  d(Q_e, Q_m) = \oint {\mathrm d} \Omega \,
 \frac{{\mathrm e}^{\i \pi (Q^T \Omega \,Q)}}{\Phi_{10}(\Omega)} \;,
\end{equation}
where $\Omega$ is the period matrix for a genus-2 Riemann surface; it
parametrizes the $\mathrm{Sp}(2)/\mathrm{U}(2)$ cosets and can be written as a 
complex, symmetric, two-by-two matrix. In the exponential factor the direct
product of the period matrix with the invariant metric of the charge lattice
(the latter is suppressed in (\ref{eq:dvvdeg})) 
is contracted with the charge vector $(\Qm,\Qe)$ comprising the 28 magnetic and
28 electric charges. This formula was conjectured based on the fact that it
generalizes the expression (\ref{eq:het-string}) for the degeneracies of
electric heterotic string states to an expression that is manifestly covariant
with respect to $S$-duality. In what follows we shall be using the
parametrization, 
\begin{equation}
  \label{eq:oq}
  \Omega = 
  \begin{pmatrix} 
    \rho & \upsilon \\ \upsilon &\sigma 
  \end{pmatrix}\,, \qquad  Q = \begin{pmatrix}\Qm \\ \Qe \end{pmatrix}\,,
\end{equation}
so that $Q^T \Omega \,Q = \rho\, \Qm^2 + \sigma\,\Qe^2  +2\, \upsilon\,\Qem$. 

The period matrix $\Omega$ transforms under $\mathrm{Sp}(2,\mathbb{Z})$
transformations, which can be written as a four-by-four matrix decomposed into
four real two-by-two blocks, $A$, $B$, $C$, and $D$ according to, 
\begin{eqnarray}
  \label{eq:SP-decomp}
\begin{pmatrix}
A&B \\C& D
\end{pmatrix}
\quad &\text{with}  &\quad
\begin{array}{l}
  A^{\mathrm T}D -C^{\mathrm T}B = D\,A^{\mathrm T} - C\,B^{\mathrm T}
  ={\bf 1}_2\,, \\
  A^{\mathrm T}C= C^{\mathrm T}A\,,\qquad B^{\mathrm T}D= D^{\mathrm
  T}B \,.
\end{array} 
\end{eqnarray}
In terms of these sub-matrices, $\Omega$ transforms as follows under
$\mathrm{Sp}(2,\mathbb{Z})$, 
\begin{equation}
  \label{eq:SP2}
   \Omega \rightarrow \Omega' = (A\,\Omega + B)\,(C \,\Omega  + D)^{-1}\,.
\end{equation}
An important related result is, $\Omega-\bar\Omega \to (\Omega\,C^{\mathrm T}
  +D^{\mathrm   T})^{-1} (\Omega-\bar\Omega)\, (C\,\bar\Omega + D)^{-1}$. 

Modular forms $\Phi_p(\Omega)$ of degree $p$ transform under the modular group
$\mathrm{Sp}(2,\mathbb{Z})$ according to 
\begin{eqnarray}
  \label{eq:Siegelform}
\Phi_p(\Omega') = \mathrm{det}(C \,\Omega + D)^p \, \Phi_p(\Omega)  \,.
\end{eqnarray}
These are holomorphic functions over the Siegel half-space, defined by
$\det(\Omega-\bar\Omega) <0$.  The modular form appearing in (\ref{eq:dvvdeg})
is the unique cusp form of degree 10. It is proportional to the square
of the Siegel cusp form $\Delta_5(\Omega)$, which is of degree 5 and has a
non-trivial multiplier system (i.e., there are extra sign factors in
(\ref{eq:Siegelform}) depending on the particular $\mathrm{Sp}(2,\mathbb{Z})$
element). The cusp form can be defined as a product over all even
theta-constants. From its behaviour under modular transformations, it follows
that it can be defined as a Fourier series with unique coefficients
(see, e.g. \cite{Gritsenko:1995xxx}), 
\begin{eqnarray}
  \label{eq:cusp-form}
  \Delta_5(\Omega)&=& \sum_{\{k,l,m\}} f(k,l,m)\, \exp\left[\i
  \pi(k\,\rho+l\,\sigma+   m\,\upsilon) \right]\,,
\end{eqnarray}
where the sum extends over $k,l,m = 1 \,\mathrm{mod}\, 2$ with $4kl-m^2>0$ and
$k,l>0$. The $f(k,l,m)$ are integral coefficients; for instance, one has
$f(1,1,1)=-f(1,1,-1)=64$. Obviously, a corresponding expansion exists for 
$\Phi_{10}$.{}\footnote{
  An alternative form of the Fourier series involves products, 
  \begin{eqnarray}
    \label{eq:product-form}
   \Phi_{10}(\Omega) = \left[\tfrac1{64} \Delta_5(\Omega)\right]^2 =
   q_\rho q_\sigma q_\upsilon 
   \prod_{\{k,l,m\}} (1- q_\rho^{\,k}\, q_\sigma^{\,l}\, 
    q_\upsilon^{\,m})^{c(kl,m)}\,,    \nonumber 
  \end{eqnarray}
where $q_\rho= \exp(2\pi\i\rho)$, $ q_\sigma = \exp(2\pi\i\sigma)$ and
$q_\upsilon = \exp(2\pi\i\upsilon)$. Here the product extends over integers
$k,l,m$, with $k,l\geq0$, or, when $k=l=0$, with $m<0$. The constants
$c(kl,m)$ depend only on $4kl-m^2$ and are 
related to the elliptic genus of $K3$; they vanish for $4kl-m^2< -1$.  
}  

The cusp form has single zeroes; one is at $\upsilon=0$ and the other ones are
in the $\mathrm{Sp}(2,\mathbb{Z})$ image of $\upsilon=0$. The zero at
$\upsilon=0$ is obvious from the relation, 
\begin{equation}
\label{|phi-zero}
  \Phi_{10}(\Omega) \approx \upsilon^2\, \eta^{24}(\rho)\, \eta^{24}(\sigma) 
\,,  
\end{equation}
which, for instance, follows from the representation of $\Phi_{10}(\Omega)$ in
terms of even theta-constants, and which we shall be using later. Here we note
in passing that $\Phi_{10}(\Omega)$ is an 
even function of $\upsilon$, as follows from (\ref{eq:Siegelform}) by applying
a transformation with $A=D=\mathrm{diag}\,(1,-1)$ and $B=C=0$. The zeroes
emerge as poles in the integrand of (\ref{eq:dvvdeg}) and therefore an 
integral over a 3-cycle that encloses such a pole will correspond
to a particular coefficient in the Fourier series for the inverse of
$\Phi_{10}$. However, the poles are located in the interior of the
Siegel half-space and not just at its boundary. Therefore the choice of the 
3-cycles in (\ref{eq:dvvdeg}) is subtle, just as the corresponding definition
of the coefficients of the Fourier  series of $(\Phi_{10})^{-1}$, as one could
be picking up extra finite residues when moving the cycle through the Siegel 
half-space. This aspect should be borne in mind when considering the
$S$-duality covariance of the expression (\ref{eq:dvvdeg}).  

Formally, the $S$-duality covariance of (\ref{eq:dvvdeg}) follows from the fact
that the effect of the transformation (\ref{eq:S-duality}) of the charges
can be compensated for by a special
$\mathrm{Sp}(2,\mathbb{Z})$ transformation on the period matrix, $\Omega\to
A\,\Omega \,A^{\mathrm{T}}$ (possibly up to integer real shifts associated 
with the sub-matrix $B$). This 
corresponds to taking $D^{-1}=A^\mathrm{T}$, and $C=0$; choosing
$A=\left(\begin{smallmatrix} a&-b \\-c&d\end{smallmatrix}\right)$, with
$ad-bc=1$, induces the required ${\mathrm{SO}}(2,1)$ transformations of
$(\rho, \sigma,\upsilon)$, 
\begin{eqnarray}
\label{eq:modparatrans}
    \rho &\to& a^2\,\rho +b^2\,\sigma -2\,ab\,\upsilon\,, \nonumber\\
  \sigma &\to& c^2 \rho + d^2 \,\sigma -2\,cd\,\upsilon\,, \nonumber\\  
  \upsilon &\to& {}-ac\, \rho -bd\,\sigma +(ad+bc) \upsilon \,.
\end{eqnarray}
The fact that the automorphic form $\Phi_{10}$ is invariant under this
subgroup of $\mathrm{Sp}(2,\mathbb{Z})$ then formally ensures the
$S$-duality covariance of (\ref{eq:dvvdeg}). A more rigorous argument along
these lines 
should in principle yield the $S$-duality invariant charge lattice, but we are
not aware of such a result in the literature. 

In the next section we will be studying the large charge limit of
(\ref{eq:dvvdeg}) by first picking out the residue from the integral over
$\upsilon$ followed by a saddle-point approximation to perform the
remaining integrals in $\rho$ and $\sigma$. Here the stationarity requirement
of the saddle-point method will implictly deal with the issue of choosing the
appropriate 3-cycles. 

\section{Asymptotic density of dyon states}
In \cite{Dijkgraaf:1996it} it is argued that, in the limit of large
charges, the leading behaviour of the degeneracy of dyon states is
determined by poles associated with the rational quadratic divisor,
\begin{equation}
  \mathcal{D} = \upsilon + \rho \sigma- \upsilon^2 = 0\,.
\label{divisorD}
\end{equation}
Subsequently, it was shown that the leading-order contribution agrees
with the macroscopic black hole entropy based on the area law. 
We expect that the subleading contributions can be extracted from the same
pole terms, up to exponentially suppressed contributions.  In order to
compute the subleading contributions, we need to know the form of the
automorphic form $\Phi_{10}(\Omega)$ in the vicinity of the divisor
(\ref{divisorD}). This can be obtained from the degeneracy limit
$\upsilon\to 0$, for which the automorphic form $\Phi_{10}(\Omega)$ has the
behaviour already indicated in (\ref{|phi-zero}),  
\begin{equation}
\label{v0}
  \frac1{\Phi_{10}(\Omega)} \longrightarrow \frac1{\upsilon^2}\,
  \frac1{\eta^{24}(\rho)\,   \eta^{24}(\sigma)}  +{\cal O}(\upsilon^0) \,.  
\end{equation}
The divisor $\upsilon=0$ is related to the divisor (\ref{divisorD})
by a $\mathrm{Sp}(2,\mathbb{Z})$ transformation given by $-B=C={\mathbf 1}_2$,
$D=0$, and $A=\left(\begin{smallmatrix} 0&1 \\1&0\end{smallmatrix}\right)$,
which yields  
\begin{equation}
\label{modtr}
  \begin{pmatrix}
    \rho & \upsilon\\[1mm] \upsilon & \sigma
  \end{pmatrix} \longrightarrow  \begin{pmatrix}
    \rho' & \upsilon'\\[1mm] \upsilon' & \sigma'
  \end{pmatrix}= \frac{1}{\mathrm{det\,\Omega}} 
  \begin{pmatrix}
    -\sigma & \upsilon + \rho\sigma -\upsilon^2 \\[1mm] \upsilon +
    \rho\sigma -\upsilon^2 & -\rho
  \end{pmatrix} \;.
\end{equation}
This transformation determines an expansion of $\Phi_{10}(\Omega)$ similar to
(\ref{v0}) as $\mathcal{D} \rightarrow 0$,
\begin{equation}
\label{phiD}
  \frac1{\Phi_{10}(\Omega)} = \frac{\det(\Omega)^{10}}{\Phi_{10}(\Omega')}
  \longrightarrow  
  \frac1{\mathcal{D}^2}\,\frac{\det(\Omega)^{12}}{\eta^{24}(\rho') \,
  \eta^{24}(\sigma')} + {\cal O}(\mathcal{D}^0)  \,. 
\end{equation}
The arguments of the Dedekind eta-functions are given by
(\ref{modtr}),
\begin{equation}
  \label{eq:r-s-prime}
  \rho^\prime = -\frac {\sigma}{\rho\sigma-\upsilon^2}\,,\qquad 
   \sigma^\prime = -\frac {\rho}{\rho\sigma-\upsilon^2}\,. 
\end{equation}
The contributions from other divisors will be exponentially
suppressed and we can now insert expression (\ref{phiD}) into
(\ref{eq:dvvdeg}) and evaluate the contour integral for $\upsilon$ around the
poles  $\upsilon_\pm= \tfrac12 \pm \Lambda$, where we have defined 
$\Lambda = \sqrt{\tfrac14+ \rho\sigma}$. Introducing 
$\gamma' = -1/\rho'$, we find that the integrand for the remaining integral
over $\rho$ and $\sigma$ becomes 
\begin{equation}
\Delta(\rho,\sigma) 
\exp\left[\i \pi \qem + \i \pi X(\rho,\sigma)
\right]\,,
\label{eq:expon}\
\end{equation}
where 
\begin{eqnarray}
X(\rho,\sigma) &=& \rho \, \qm^2 + \sigma \, \qe^2 \pm 2\, \Lambda
\, \qem 
+ \frac{12 \alpha}{\i \pi} \log \sigma 
- \frac{24 \alpha}{\i \pi} \log \eta (\sigma')
- \frac{24 \alpha}{\i \pi} \log \eta (\gamma') \;,
\nonumber\\[1ex]
\Delta(\rho,\sigma) & = & \frac{1}{4\Lambda^2} \left[ 2 \pi \i \,\qem \mp
  \frac{1}{\Lambda} + 48 \alpha \, \left(  \sigma'  
\frac{ \d \log \eta (\sigma')}{\d  \sigma'}
-\gamma' \frac{ \d \log \eta (\gamma')}{\d \gamma'}
 \right)\right]\;.
\label{XZ}  
\end{eqnarray}
The term $\i\pi\Qem$ in (\ref{eq:expon}) represents an overall sign factor as
$\Qem$ is expected to take integer values, so that we will drop it in the
following. Furthermore, we have introduced the parameter $\alpha$, which is
given by $\alpha=1$,  in order to keep track of the terms
coming from the Dedekind eta-functions.  This parameter is the counterpart
to $-64 c_1 $ in (\ref{eq:nonholoentropy}). In order to arrive at this
result, we have  used the modular properties of the Dedekind eta-function
to express $\eta^{24} (\rho')$ in terms of $\eta^{24} ( - 1/\rho')$, which
gives rise to the term $12 \alpha \log \sigma$ in $\i \pi X(\rho,\sigma)$. In
doing so the factors $\det(\Omega)$ cancel. 

It is instructive to express the integrand solely in terms $\sigma'$
and $\gamma'$.  To this extent we note the following identities valid
on the divisor (\ref{divisorD}), 
\begin{eqnarray}
\label{valueSigma}
v_\pm = \tfrac{1}{2} \pm \Lambda =  \frac{\gamma'}{\sigma'+\gamma'} \;,\quad
\rho = \frac{\sigma'\gamma'}{\sigma'+\gamma'}\;,\quad \sigma =  -
  \frac{1}{\sigma' + \gamma'}\,.
\end{eqnarray}
Substituting these expressions in (\ref{XZ}), 
the exponent takes the suggestive form 
\begin{eqnarray}
\label{eq:niceX}
 \i \pi  X(\sigma, \rho)& =& -    \pi \, \left[\frac{\qe^2  +
    (\sigma'-\gamma') \qem- 
    \sigma'\gamma' \qm^2}{-\i (\sigma'+\gamma')}\right] - 2 \alpha \log
\left[(\sigma'+\gamma')^6  
  \eta(\sigma')^{12} \eta(\gamma')^{12}\right] \,,\nonumber \\
 && {~}
\end{eqnarray}
which holds for both poles at $v=v_\pm$. At this point we observe a remarkable 
fact: if one identifies,  
\begin{equation}
\label{eq:identif}
  \sigma' =\i \bar S\;,\qquad \gamma' = \i S\,, 
\end{equation}
in the expression (\ref{eq:niceX}), it precisely coincides with the
macroscopic entropy formula (\ref{eq:nonholoentropy}) presented in
section~1. Also the arguments of the Dedekind eta-functions match: the
functions $\eta(S)$ and $\eta(\bar S)$ that appear in 
(\ref{eq:nonholostab}) 
and (\ref{eq:nonholoentropy}) are functions of the argument $q =
\e^{-2 \pi S}$ and $q= \e^{-2 \pi \bar S}$, respectively, while in the
microscopic approach $\eta(\sigma')$ and $\eta(\gamma')$ are functions
of $q = \e^{2 \pi \i \sigma'}$ and $q = \e^{2\pi \i \gamma'}$,
respectively.

Before proceeding let us first consider some consequences of this surprising
match. First of all, the identification implies that $\sigma^\prime$ is
equal to minus the complex conjugate of $\gamma^\prime$, and therefore the
expression (\ref{eq:niceX}) is real. Secondly, one may wonder what the
consequences are for $S$-duality. To investigate this, let us determine the
$S$-duality transformations on $\rho$, $\sigma$ and $\upsilon_\pm$ as induced
by the $S$-duality variations of $S$, through (\ref{valueSigma}). A simple
calculation yields the following result,  
\begin{eqnarray}
  \label{eq:induced-S}
    \rho &\to& a^2\,\rho +b^2\,\sigma -2\,ab\,\upsilon_\pm + ab\,, \nonumber\\
  \sigma &\to& c^2 \rho + d^2 \,\sigma -2\,cd\,\upsilon_\pm + cd \,,
  \nonumber\\   
  \upsilon_\pm &\to& {}-ac\, \rho -bd\,\sigma +(ad+bc) \upsilon_\pm -bc \,.
\end{eqnarray}
Hence, these transformations coincide with the transformations
(\ref{eq:modparatrans}), up to translations by integers. In fact, they
constitute the subgroup of $\mathrm{Sp}(2,\mathbb{Z})$ that 
leaves the divisor (\ref{divisorD}) invariant. This explains why they apply
irrespective of which pole one chooses. Based on these
observations, the identification (\ref{eq:identif}) seems to be a very
sensible one indeed. 

The remaining two integrals associated with the 3-cycle will be carried out in
a saddle-point approximation. 
Note, however, that the integrand (\ref{eq:expon}) also contains a
contribution from the factor $\Delta(\rho,\sigma)$. Of course, both
$X(\rho,\sigma)$ and $\Delta(\rho,\sigma)$ will enter in the saddle-point
evaluation of the integral and in principle, 
one should determine the saddle-point values of $\rho$ and $\sigma$ from the
extremality conditions of the complete integrand. Nevertheless, we will
treat the two pieces $X(\rho,\sigma)$ and $\Delta(\rho,\sigma)$ of the
integrand (\ref{eq:expon}) separately. While $\Delta(\rho,\sigma)$
does contribute (logarithmically) to the saddle-point value of the
integrand, we initially neglect $\Delta(\rho,\sigma)$ when determining the
saddle-point. Hence we will be expanding the integrand around an
approximate extremal point and after performing the integrations there will be
additional contributions involving derivatives of $\log(\Delta(\rho,\sigma))$.
As we shall see, these contributions are suppressed by inverse powers of
derivatives of $X(\rho,\sigma)$ and the approximation is reliable because the
derivatives of $X(\rho,\sigma)$ contain terms proportional to the
charges, unlike the derivatives of $\log(\Delta(\rho,\sigma))$.  

The saddle-point equations derived from $\exp(\i \pi X(\rho,\sigma))$
are given by
\begin{equation}
\i \pi
  \partial_\rho X = \i \pi \qm^2
+ 2 \,\i \,\pi \qem\, \frac{1}{\sigma'-\gamma'}
+24 \alpha\, \frac{\sigma'+\gamma'}{\sigma'-\gamma'}
  \left[ \frac{ \d \log \eta (\sigma')}{\d \sigma'}
- \frac{ \d \log \eta (\gamma')}{\d \gamma'}
\right]        =0 \;,
\label{eq:critX1}
\end{equation}
and
\begin{multline}
\qquad \quad \i \pi \partial_\sigma X  = \i \pi \qe^2 
-2\, \i\, \pi(\qem)\, \frac{  \sigma' \gamma'}{\sigma'-\gamma'} 
 \\  -12\alpha(\sigma'+\gamma')
-24 \alpha \, \frac{\sigma'+\gamma'}{\sigma'-\gamma'}
  \left[ \sigma'^2 \,
  \frac{ \d \log \eta (\sigma')}{\d \sigma'}-
\gamma'^2 \frac{ \d \log \eta (\gamma')}{\d \gamma'}
\right] = 0 \;.\quad 
\label{eq:critX2}
\end{multline}
Note that $\i \pi X(\rho,\sigma)$ and the saddle-point equations
(\ref{eq:critX1}) and (\ref{eq:critX2}) derived from it hold irrespective of
the pole at $\upsilon_\pm$ that has been selected in the initial contour
integral. The only dependence resides in the relation between
$\sigma',\gamma'$ and $\rho,\sigma$, specified by (\ref{valueSigma}). 
But since the saddle-point equations are identical for both  poles, the choice
of the pole is irrelevant. 

With the identification (\ref{eq:identif}) the saddle-point equations
(\ref{eq:critX1}) and (\ref{eq:critX2}) precisely coincide with the
non-holomorphic attractor equations (\ref{eq:nonholostab}).  
We conclude that a solution to the saddle-point equations is provided by
\begin{equation}
\label{eq:solu}
\sigma'|_0= \i \bar S\;,\quad \gamma'|_0=  \i S  \;,
\end{equation}
where $S$ is subject to (\ref{eq:nonholostab}).For this solution, $\i \pi X$
given in (\ref{eq:niceX}) is precisely equal to the macroscopic black hole
entropy! Although it is conceivable that the condition $\gamma' = -\bar
\sigma'$, which is implied by (\ref{eq:identif}), can be relaxed and
other saddle-point values than those dictated by the attractor
equations (\ref{eq:nonholostab}) can be found, we believe this to be
unlikely. Firstly, for the leading, $\alpha$-independent terms in
(\ref{eq:niceX}) to be real, $\sigma'+\gamma'$ must be imaginary and
$\sigma'-\gamma'$ real. This implies that $\gamma' = -\bar \sigma'$.
Secondly, we prove (\ref{eq:solu}) by direct calculation in the limit where
$\sigma'$ and $\gamma'$ have large and positive imaginary 
parts. In this limit, we can expand the Dedekind eta-functions and solve the  
saddle-point equations for $\sigma'-\gamma'$ and $\sigma'+\gamma'$.
The equations (\ref{eq:critX1}) and (\ref{eq:critX2}) simplify and one
finds
\begin{equation}
   \sigma' - \gamma' = -\frac{2\,\qem}{\qm^2}\,.
\end{equation}
Inserting this expression into (\ref{eq:critX2}) we find the following
equation for $\sigma'+\gamma'$,
\begin{equation}
\label{eq:quadsg}
  -\tfrac{1}{4}{\qm^2} (\qm^2 - 8 \alpha)(\sigma'+\gamma')^2 + \frac{12
    \alpha}{\i\pi}{\qm^2} (\sigma'+\gamma') = \qe^2\,\qm^2 - (\qem)^2 \,.
\end{equation}
This is a quadratic equation which uniquely determines the value of
$\sigma'+\gamma'$  in terms of the charges. A comparison of (\ref{eq:quadsg})
with (\ref{eq:re-S}) shows that the saddle-point values for
$-\i(\sigma'+\gamma')$ and $\i(\sigma' - \gamma')$ are precisely the
attractor values for the real and imaginary parts of the dilaton
$S$. Hence, we are necessarily led to $\sigma' - \gamma' = -\i (S-\bar S)$ and
$\sigma' + \gamma'= \i (S+\bar S)$ in accord with the identification
(\ref{eq:identif}).

Let us return to saddle-point approximation for the remaining integrals. 
As stressed above, we now have to include the factor $\Delta(\rho,\sigma)$,
which has the following form,
\begin{eqnarray}
  \label{eq:Delta}
&&\log \Delta(\rho,\sigma) = \log\left[ 
\frac{2\i \pi  \,\qem\,(\sigma'+\gamma')^2}{(\sigma'- \gamma')^2}\right] 
\nonumber \\[1ex]
&&\quad
+ \log\left[1+  \frac1{\i \pi \,\qem}\left(\sigma'  
\frac{ \d}{\d\sigma'}-\gamma' \frac{\d}{\d \gamma'}\right)
\left(\log\left[\sigma'- \gamma'\right] + 2\, \alpha \,   
\log \left[\eta^{12} (\sigma')\, \eta^{12}(\gamma')\right]  \right)\right]\;.
\nonumber \\
&& {~} 
\end{eqnarray}
Furthermore we need the expression for the matrix of the second derivatives of
$\i\pi  X(\rho,\sigma)$, which takes the form
\begin{equation}
  \label{eq:d2-X}
  \frac{\partial^2 \left(\i\pi\,X(\rho,\sigma)\right)} 
  {\partial^2(\rho,\sigma)} = {}  -\frac{2\i\pi\Qem
  (\sigma'+\gamma')}{(\sigma'-\gamma')^3}  
\begin{pmatrix} - 2&\sigma^{\prime 2}+\gamma^{\prime 2} \\[2mm]
 \sigma^{\prime 2}+\gamma^{\prime 2} & -2\,\sigma^{\prime 2}\, 
  \gamma^{\prime 2} 
\end{pmatrix} + \cdots \;,
\end{equation}
where we suppressed terms that do not depend explicitly on the charges. 

With these expressions we can complete the saddle-point approximation. First, 
we note that the derivatives of $\log(\Delta(\rho,\sigma))$ are of order
$(\Qem)^0$, whereas the leading term of the matrix of second-order derivatives
of $\i\pi X(\rho,\sigma)$ is proportional to $\Qem$. Furthermore, we have to
perform a  
two-dimensional integral over the real values of $\rho$ and $\sigma$. 
Consequently, the result of the saddle-point approximation yields
\begin{equation}
\label{eq:saddlepointapprox}
  \mathcal{S}_{\text{micro}} =  \left. \i \pi X\right|_0 +
  \left.\log \Delta\right|_0 -
  \left.\tfrac{1}{2}\log \det 
  \frac{\partial^2 (\i \pi X(\rho,\sigma))}{\partial^2(\rho,\sigma)} \right|_0
  \,, 
\end{equation}
up to terms that behave inversely proportional to $\Qem$. 
Note, however, that the second and third term depend only on $\Qem$, whereas
the first term depends on all three combinations $\Qe^2$, $\Qm^2$ and $\Qem$
and it is known to be invariant under $S$-duality. Therefore, one expects that
the contributions from the second and third term will cancel. This is indeed
the case: computing the last term of (\ref{eq:saddlepointapprox}), we find
that it is given by 
\begin{equation}
  \left.-\tfrac{1}{2} \log \det \frac{\partial^2 (\i \pi
      X(\rho,\sigma))}{\partial^2(\rho,\sigma)}\right|_0  = 
- \log \left[\frac{2\pi\i \qem
      \,(\sigma'+\gamma')^2}{(\sigma'-\gamma')^2}\right]  +
   {\cal O}(1/ \qem)\,. 
\end{equation}
This term cancels exactly against the first term in
(\ref{eq:Delta}). Likewise, the attractor equations, which correspond to the
saddle-point equations (\ref{eq:critX1}) and (\ref{eq:critX2}), are only
modified by terms that are inversely proportional to $\Qem$. 

We have thus verified that $d(\Qe,\Qm)$ defined in (\ref{eq:dvvdeg}) is equal
to $\exp(\mathcal{S}_{\mathrm{macro}})$ for large charges with nonvanishing
$\Qem$. Here $\mathcal{S}_{\mathrm{macro}}$ represents the macroscopic entropy
given in (\ref{eq:nonholoentropy}), subject to the attractor equations
(\ref{eq:nonholostab}). We expect that the same result can be established for
the case $\Qem=0$. Consequently, we have shown 
that the dyon degeneracy formula (\ref{eq:dvvdeg}) leads precisely to the
results of \cite{LopesCardoso:1999ur} summarized in equations
(\ref{eq:nonholostab}) and (\ref{eq:nonholoentropy}). 

\section{Conclusions and Outlook}
In the previous section we have established that the formula
(\ref{eq:nonholoentropy}) for the macroscopic entropy 
is in agreement with the microscopic degeneracy of dyons proposed in
\cite{Dijkgraaf:1996it}.
The former includes non-holomorphic terms which are crucial 
for obtaining an $S$-duality invariant result. Here we have demonstrated that
these non-holomorphic terms are precisely captured by the microscopic
counting. 

Let us first discuss the structure of the non-holomorphic corrections, also in
the light of the observation in \cite{Ooguri:2004zv} that the formula for the
black hole entropy formula can be reinterpreted as the Legendre transform of
the black hole free energy. The entropy formula of \cite{LopesCardoso:1998wt}
is based on a supersymmetric  Wilsonian effective action, which for $N=2$
supergravity 
is encoded in a homogeneous, holomorphic function of projectively defined
quantities. In the formalism of \cite{LopesCardoso:1998wt} a convenient set of
complex variables was found, denoted by $Y^I$ and $\Upsilon$. The variable 
$\Upsilon$ is associated with an extra chiral supermultiplet related to
the Weyl multiplet of conformal supergravity and its presence in the
holomorphic function gives rise to terms in the effective action proportional
to the square of the Riemann  
curvature. The relevant function for the heterotic case takes the following
form \cite{LopesCardoso:1999ur}, 
\begin{equation}
  \label{eq:F}
  F(Y,\Upsilon) = - \frac{Y^1\,Y^a\eta_{ab}Y^b}{Y^0} +  F^{(1)}(Y^1/Y^0) 
  \,\Upsilon \,,
\end{equation}
where $F^{(1)}$ is some function of the dilaton field $S=-\i Y^1/Y^0$.  
The entropy formula for generic homogeneous and holomorphic functions $F$ can
be written as 
\begin{equation}
  \label{eq:holo-entropy}
  \mathcal{S}_{\text{macro}} = \pi\left[ p^IF_I(Y,\Upsilon) - q_IY^I -
  256\,{\mathrm{Im}}(F_\Upsilon(Y,\Upsilon)) \right]_{\Upsilon=-64} \,,
\end{equation}
where the value of $-64$ represents the value of $\Upsilon$ taken at the
horizon. Likewise, the electric and magnetic charges $q_I$ and $p^I$,
respectively, determine the horizon values of the $Y^I$
according to the attractor equations (using the horizon value $\Upsilon=-64$),
\begin{equation}
  \label{eq:hol-attractor}
  Y^I- \bar Y^I= \i p^I\,,\qquad F_I(Y,\Upsilon)-\bar F_I(\bar Y,\bar\Upsilon)
  = \i q_I\,,
\end{equation}
where $F_I=\partial F/\partial Y^I$ and $F_\Upsilon= \partial
F/\partial\Upsilon$. The first two terms
in~(\ref{eq:holo-entropy}), which are real by virtue
of~(\ref{eq:hol-attractor}), represent one-fourth of the black hole area,
while the last term proportional to $F_\Upsilon$ represents the deviation of
the area law.  

In \cite{LopesCardoso:1999ur} the function $F^{(1)}(S)$ in (\ref{eq:F}) was
determined by requiring target-space duality and $S$-duality invariance. For
achieving consistency with string perturbation theory one must introduce
non-holomorphic terms. The approach that was followed for deriving these terms
was somewhat ad hoc. It was first assumed that the attractor equations
(\ref{eq:hol-attractor}) for $Y^I$ and $F_I$ still hold, but that these
quantities contain non-holomorphic terms, corresponding to 
\begin{equation}
  \label{eq:F1}
  F^{(1)} (S,\bar S) = - \frac{\i c_1}{\pi} \log\left[ (S+\bar S)^6
  \eta^{12}(S)\right] \;. 
\end{equation}
However, when substituting this modification into the entropy formula
(\ref{eq:holo-entropy}) it did not produce an $S$-duality
invariant result and one had to introduce yet another term equal to $-128\,
c_1 \log(S+\bar S)^6$. 
The combined result of these modifications is concisely summarized in
(\ref{eq:nonholostab}) and (\ref{eq:nonholoentropy}).
As we have already explained in the introduction, the presence of
non-holomorphic corrections is to be expected in view of the fact that the
Wilsonian action, which is based on holomorphicity, does in general not reflect
the symmetries of the theory. 

In \cite{Ooguri:2004zv} the result of the holomorphic case was reformulated in
terms of a real function and it is of interest to see how the non-holomorphic
terms will manifest themselves in that formulation. This is a priori not
completely obvious as both the holomorphicity and the homogeneity of the
function $F$ were used in the derivation. In \cite{Ooguri:2004zv} the $Y^I$
were expressed 
in terms of the magnetic charges $p^I$ and (real) electrostatic potentials
$\phi^I$ at the horizon, 
\begin{equation}
\label{eq:electro-phi} 
  Y^I = \frac{\phi^I}{2\pi} + \frac{\i p^I}{2} \,,
\end{equation}
so that the first set of attractor equations~(\ref{eq:hol-attractor}) is 
already incorporated. The remaining attractor equations
(\ref{eq:hol-attractor}) and the entropy (\ref{eq:holo-entropy}) can then be
written as follows,  
\begin{eqnarray}
  \label{eq:real-F}
  q_I&=&   \frac{\partial\mathcal{F}(\phi,p)}{\partial \phi^I} \,, \nonumber\\ 
  \mathcal{S}_{\text{macro}}(p,q) &=& \mathcal{F}(\phi,p) - \phi^I
  \frac{\partial\mathcal{F}(\phi,p)}{\partial \phi^I} \,,
\end{eqnarray}
where the real function $\mathcal{F}(\phi,p)$ is defined by 
\begin{equation}
  \label{eq:def-real-F}
  \mathcal{F}(\phi,p) = 4\pi \,\mathrm{Im} [\,F(Y,\Upsilon)]_{\Upsilon=-64} \,.
\end{equation}
This function is related to the entropy by a Legendre transformation and
therefore it will be called the thermodynamic free energy. 
The homogeneity of the function $F(Y,\Upsilon)$ was crucial for deriving 
(\ref{eq:real-F}) and (\ref{eq:def-real-F}) and use was
made of the corresponding identity 
\begin{equation}
  \label{eq:homogeneity}
   Y^I F_I(Y,\Upsilon) + 2\,\Upsilon F_\Upsilon(Y,\Upsilon) = 2\,F(Y,\Upsilon)
   \,.
\end{equation}

To study the effect of (possibly non-holomorphic) corrections, let us simply
modify $\mathcal{F}$ according to 
\begin{equation}
  \label{eq:macrofreeenergy}
  {\widehat {\cal F}}(\phi,p) = 4 \pi \,\mathrm{Im}[\, F(Y,\Upsilon)] +4\pi \,
  \Omega(Y,\bar Y,\Upsilon,\bar\Upsilon) \,,
\end{equation}
where for the moment we refrain from setting $\Upsilon=-64$. Here
$\Omega$ denotes a real function. We then assume that
(\ref{eq:real-F}) remains valid with $\mathcal{F}(\phi,p)$ replaced by
$\widehat{\mathcal{F}}(\phi,p)$, where (\ref{eq:electro-phi}) still defines 
the magnetic charges and electrostatic potentials. The electric charges are
thus defined by  
\begin{equation}
  \label{eq:hat-q}
  \frac{\partial}{\partial Y^I}(F +2 \i \, \Omega) -  \frac{\partial}{\partial 
  \bar Y^I}(\bar F -2 \i \, \Omega) = \i \,\frac{\partial{\widehat {\cal F}}}
  {\partial \phi^I} = \i q_I\,,
\end{equation}
while the entropy can now be written as
\begin{eqnarray}
  \label{eq:Delta-entropy}
  \mathcal{S}_{\text{macro}} &=& \pi \Big[ p^I \,\frac{\partial}{\partial
    Y^I}(F +2 \i \, \Omega) - q_I\, Y^I - 2\i\,   
    (\Upsilon\, F_\Upsilon -\bar\Upsilon\,\bar F_{\bar\Upsilon}) \nonumber
    \\[1ex] 
&& \qquad + 2\,\Big( 2 - Y^I\frac{\partial}{\partial Y^I} -\bar
    Y^I\frac{\partial}{\partial \bar Y^I}\Big)\,\Omega\,  \Big ] \,.
\end{eqnarray}
In deriving this result we used (\ref{eq:hat-q}) and the homogeneity and
holomorphicity of $F(Y,\Upsilon)$.  

Furthermore, if we assume that $\Omega$ is a homogeneous function of degree 2,
so that 
\begin{equation}
  \label{eq:hom-Omega}
  2\,\Omega - 
    Y^I\frac{\partial\Omega}{\partial Y^I} -\bar
    Y^I\frac{\partial\Omega}{\partial \bar Y^I} =  2\,\Upsilon
    \frac{\partial\Omega} {\partial \Upsilon} +2\,\bar\Upsilon
    \frac{\partial\Omega} {\partial \bar\Upsilon} \,, 
\end{equation}
we can make direct contact with (\ref{eq:holo-entropy}) via 
the substitutions
\begin{eqnarray}
  \label{eq:subst-Omega}
  F_I(Y,\Upsilon) &\longrightarrow&  F_I(Y,\Upsilon) + 2\i\,
  \frac{\partial \Omega(Y,\bar Y,\Upsilon,\bar\Upsilon)}{\partial Y^I}
  \,,\nonumber \\[1ex]
  F_\Upsilon(Y,\Upsilon) &\longrightarrow& F_\Upsilon(Y,\Upsilon) + 2\i\,
  \frac{\partial \Omega(Y,\bar Y,\Upsilon,\bar\Upsilon)}{\partial \Upsilon}
  \,. 
\end{eqnarray}

So far we have not made any assumption about the holomorphic behaviour of
$\Omega$. Clearly, when $\Omega$ can be written as the imaginary part of a
holomorphic function $\Omega(Y,\Upsilon)$, then we recover the previous result
with the function 
$F(Y,\Upsilon)$ replaced by $F(Y,\Upsilon) + \Omega(Y,\Upsilon)$. 
However, the result can equally well be applied to more general
corrections. For instance, assuming that $F(Y,\Upsilon)$ is just equal to the 
classical result, characterized by the $\Upsilon$-independent first term in
(\ref{eq:F}), we find the complete result (\ref{eq:nonholostab}) and
(\ref{eq:nonholoentropy}) by substituting the following expression for
$\Omega$, 
\begin{equation}
   \label{eq:Omega}
   \Omega(Y,\bar Y,\Upsilon,\bar \Upsilon) = - \frac{c_1}{2\pi}\left[\Upsilon
   \log \eta^{12}(S) + \bar\Upsilon \log \eta^{12}(\bar S)
   +\tfrac1{2}(\Upsilon+ \bar \Upsilon) \log (S+\bar S)^6 \right] \,.
\end{equation}
At this point we are thus able to give the corresponding expression for
$\widehat{\mathcal{F}}(\phi,p)$, where now we have substituted $\Upsilon=-64$,
\begin{eqnarray}
  \label{eq:F-hat-het}
  \widehat{\mathcal{F}}(\phi,p) &=& (S+\bar S)\,
  \eta_{ab}\,\Big(\frac{\pi\, p^ap^b} {2}
  -   \frac{\phi^a\phi^b}{2\pi} \Big) - \i (S-\bar S)
  \,\eta_{ab}\,p^a\phi^b\nonumber \\[1ex]
  &&
  + 128\,c_1 \log \left[(S+\bar S)^6 \,\vert\eta(S)\vert^{24} \right]
  \,, 
\end{eqnarray}
where
\begin{equation}
  \label{eq:S-value}
  S= -\i\, \frac{\phi^1 + \i\pi p^1}{\phi^0 + \i\pi p^0} \,.
\end{equation}
Thus we see that the real function (\ref{eq:F-hat-het}) associated to the
macroscopic entropy (\ref{eq:nonholoentropy}) contains non-holomorphic terms,
and that these appear in precisely the combination that corresponds to the
$S$-duality invariant physical coupling function of the $R^2$-terms 
\cite{Harvey:1996ir}. 
It would be interesting to compare this real function with the one obtained by
performing the Laplace transform of the microscopic degeneracy formula
(\ref{eq:dvvdeg}). The non-holomorphic corrections in (\ref{eq:F-hat-het}) are
presumably related to the difference between a Legendre and a Laplace
transform and associated to integrating out the moduli fields $\phi^a$ of the 
$\mathrm{SO}(6,22)$ coset. 

Finally, we comment briefly on the case of purely electric black holes. These
configurations constitute $1/2$-BPS states and have recently been
reanalyzed in \cite{Dabholkar:2004yr,Sen:2004dp}, both from the
microscopic and macroscopic perspective. We would like to point out
that it seems difficult to match the microscopic and the macroscopic
results at the non-perturbative level. To illustrate this, let us
compare the attractor equations (\ref{eq:nonholostab}) for purely electric
black holes (at the non-perturbative level, the dilaton must be real in this
case), 
\begin{equation}
  \i \pi \Qe^2 =  24 \,\i\, S - 96 \, S^2 \, \frac{\d  \log \eta(\i S)}{\d( \i
    S)}\,,
\end{equation}
with the saddle-point equation for the variable $\sigma$ derived from
the degeneracy formula (\ref{eq:het-string}),
\begin{equation}
  \i \pi \qe^2 = 12  \gamma + 24\,\gamma^2
  \frac{\d \log \eta(\gamma)}{\d \gamma}\,,
\end{equation}
where we have redefined $\gamma = -1/\sigma$. In order to facilitate the
comparison, we use that the $\eta(\tau)$ in this paragraph is always
understood as a power series in  $q = \exp(2 \pi \i \tau)$. At 
the perturbative level where $S$ is large, both equations agree upon an
identification $\gamma= 2\, \i\, S $, which is presumably related to an
observation made in \cite{Dabholkar:2004yr} about the dilaton 
normalization. Contrary to the dyonic case, this identification 
fails at the non-perturbative level, since matching the arguments of
the Dedekind eta-functions would require $\gamma = \i S$. Furthermore,
there is no way to achieve a non-perturbative matching by rescaling the
charges~$\Qe^2$ and/or the dilaton~$S$. The reason for this mismatch is not
known to us at present. Note that in this case the limit of
large charges is taken in the sublattice characterized
by $\Qe^2\, \Qm^2 -(\Qem)^2=0$, which is unrelated to the sublattice of the
dyonic configurations.   

We close with the following observation. As shown in this paper the attractor
equations (\ref{eq:nonholostab}) are identical to the saddle-point equations
which result from 
the stationarity of the quantity $X$ given in (\ref{eq:niceX}). This suggests
to view the attractor equations (\ref{eq:nonholostab}) as the conditions that
ensure that the  expression (\ref{eq:nonholoentropy}) for the macroscopic
entropy,  written as a function of the dilaton field $S$,
takes an extremal value at the horizon. The significance of this result is not
entirely clear to us, as it depends on the way in which we have written the
entropy formula. Namely, by using the attractor equations we could write the
expression (\ref{eq:nonholoentropy}) into an alternative form which takes the
same value at the horizon, but for which this stationarity condition would not
hold.  

\subsection*{Acknowledgements}
We wish to thank R.~Bruggeman, G.~Curio, K.~Haberland, 
P.~Mayr, and S.~Stieberger for valuable discussions. This work is partly
supported by EU contract MRTN-CT-2004-005104. The
work of T.M. is supported by the DFG ``Schwerpunktprogramm String\-theorie''.  
\providecommand{\href}[2]{#2}
\begingroup\raggedright\endgroup

\end{document}